\documentclass[12pt]{article}
\usepackage[utf8]{inputenc}

\usepackage[left=1in,right=1in,top=1in,bottom=1in]{geometry}
\usepackage{booktabs} 
\usepackage{authblk} 
\usepackage{graphicx} 
\usepackage{comment}
\usepackage{breakcites}
\usepackage[bookmarks=false]{hyperref}
\usepackage{amsmath}
\usepackage{amssymb}
\usepackage[dvipsnames]{xcolor}
\usepackage{lineno}

\title{Organized crime behavior of shell-company networks in procurement: prevention insights for policy and reform}

\author[1]{}
\author[1]{J. R. Nicolás-Carlock\footnote{Correspondence: nico.carlock@gmail.com, il2402@columbia.edu}}
\author[2]{I. Luna-Pla$^*$}
\affil[1]{\normalsize Institute of Physics, National Autonomous University of Mexico, Mexico}
\affil[2]{\normalsize Institute of Latin American Studies, Columbia University }
\date{}

\begin{document}

\maketitle


\begin{abstract}

In recent years, the analysis of economic crime and corruption in procurement has benefited from integrative studies that acknowledge the interconnected nature of the procurement ecosystem. Following this line of research, we present a networks approach for the analysis of shell-companies operations in procurement that makes use of contracting and ownership data under one framework to gain knowledge about the organized crime behavior that emerges in this setting. In this approach, ownership and management data are used to identify connected components in shell-company networks that, together with the contracting data, allows to develop an alternative representation of the traditional buyer-supplier network: the \textsl{module-component} bipartite network, where the modules are groups of buyers and the connected components are groups of suppliers. This is applied to two documented cases of procurement corruption in Mexico characterized by the involvement of large groups of shell-companies in the misappropriation of millions of dollars across many sectors. We quantify the economic impact of single versus connected shell-companies operations. In addition, we incorporate metrics for the diversity of operations and favoritism levels. This paper builds into the quantitative organized crime in the private sector studies and contributes by proposing a networks approach for preventing fraud and understanding the need for legal reforms.

\end{abstract}

\section*{Introduction}

The use of companies for criminal and illicit practices by power elites is a well-documented phenomenon across many nations and a longstanding focus of research in the fields of money laundering and economic crime studies \cite{VanDuyne2006, deWillebois2011, fazekas2016comprehensive, Albanese2018, Campbell2018}. In particular, a context in which companies become the most important vehicle to extract and hide the proceeds of economic crime and corruption lies at the intersection of public and private sectors, namely, public procurement \cite{oecd2016preventing, fazekas2016comprehensive, Fazekas2017}. Fraud and corruption in public procurement not only result in significant financial losses for governments but also hinder human development and the rule of law by compromising public services, infrastructure, and the overall functioning of government institutions \cite{bajpai2020, oecd2016preventing}. Procurement markets are complex systems regulated by administrative processes of different levels of monopoly, discretion, and transparency, where a great number of public and private institutions interact in contexts where legal and illegal activities can intertwine. All stages of the contracting process are prone to manipulation and abuse, with risks of embezzlement, bribery, conflict of interest, fraud, or conspiracy \cite{deases2004developing, schultz2008corruption, oecd2016preventing, fazekas2020uncovering}. As such, procurement markets tend to create environments that favor illicit activities by political, economic, and criminal elites \cite{santino2022mafia, canonico2021criminal}, and where the identification of organized crime, grand corruption schemes, or the criminal groups behind those operations is not simple \cite{conley2016detecting, Campbell2018, ReuterandPaoli2020}.

Research in organized crime has had the challenge to define, conceptually and theoretically, the illegal markets diversity and the way they shape the environment as it impacts the organization type \cite{ReuterandTonry2020, mclaughlin2010}. Nevertheless, in recent years, the study of economic crime and corruption in procurement has benefited from integrative studies that acknowledge the interconnected nature of the procurement ecosystem \cite{lyra2022fraud, corruption-networks2021, Kertesz2020, Luna-Pla2020}. For example, the use of network science and open procurement data for the modeling of the contracting relations among government institutions (buyers) and companies (suppliers) as bipartite networks has helped to measure the levels of corruption risk due to concentrations of single-bid contracting rates \cite{fazekas2020, Wachs2020}. Another important approach has dealt with the characterization of firm-firm co-bidding networks in order to detect economic cartels and their fraudulent activities in public tenders, such as collusion or cooperation \cite{Wachs2019, lyra2021}. Furthermore, it is well-known that when companies are not independent among each other but are interrelated due to shared ownership or management they pose a higher risk of fraud, collusion and market manipulation \cite{nicolas2021, velasco2021decision, jancsics2017offshoring, vitali2011, Fazekas2017}. Therefore, comprehending the interconnected operations of buyers and suppliers at all stages of the procurement process, from tenders and awarded contracts to ownership relations and other informal networks ties \cite{Costa2021} is a crucial aspect in identifying irregular activities and preventing illicit and criminal operations in these complex environments \cite{lyra2022fraud}.

This paper builds into the quantitative economic organized crime studies, by presenting a networks approach for the analysis of corruption in procurement that makes use of contracting and ownership data under one framework. Our contribution centers in proposing an analytical method for preventing fraud and corruption that quantifies the economic impact of single versus connected shell-companies operations, including metrics for the diversity of operations and favoritism levels. In this approach, ownership and management data are used to identify the connected components in one-mode shell-company networks that, together with the contracting data, allows to develop an alternative representation of the traditional buyer-supplier network: the \textsl{module-component} bipartite network, where the modules are groups of buyers and the connected components are groups of suppliers. As such, the module-component representation allows the identification of different regions of the market according to the shell-companies contracting patterns. We apply this approach to two recent documented cases of corruption in procurement characterized by the operations of large groups of shell-companies in the misappropriation of millions of dollars across multiple sectors in the Mexican states of Puebla and Guanajuato. Finally, we argue on the practical implications of our findings to the definition of economic organized crime forms, the measurement of emerging patterns, and the prevention and punishment insights that can possibly impact legal frameworks. Network analysis has the potential to identify criminal typologies that indicate higher economic and political risks \cite{fatf2018, Zumaya2021, fazekas2020, falcon2022practices}, as well as to find gaps in law enforcement or misconceived problems in legislation that hinder the control of criminal networks.

\section*{Cases \& Data}

\subsection*{Cases}

The two corruption cases chosen for this study were independently documented by investigative journalism groups and local government agencies in the Mexican states of Puebla and Guanajuato. These cases were selected due to the data availability and general similarities: in both, state and municipal government institutions awarded contracts to companies that were all prosecuted and officially enlisted as shell-companies by Mexican fiscal authorities after their investigations. Among the many causes found by authorities for such classification, there was: failure to comply with the contracts by not delivering public goods agreed upon, performing product substitution, simulating operations, and/or providing inauthentic or missing information. Here, a shell company is defined as a legal person lacking substantial assets, operations, or personnel structure, and used for illicit purposes, typically oriented to conceal beneficial ownership \cite{deWillebois2011}. According to the Mexican law, shell companies are incorporated in a public registry of tax evasion when there is a missing or false address, inauthentic documentation, lack of assets, simulated operations, or when they issue fiscal invoices to feign operations. Additionally, both cases include alleged fraud activities, money laundering, and tax evasion, often colluded and in conspiracy with medium and high-level public officials. 

\textsl{Case 1 – Puebla.} This investigation was developed by the investigative journalism group Datamos in coordination with the International Center for Journalists and Connectas\footnote{Valencia R, Velázquez M. Puebla, fábrica de empresas fantasma. Puebla: Datamos (2020).}. The data collection was made by freedom of information requests and from official websites (\textsl{Plataforma Nacional de Transparencia} and \textsl{Compranet}). This investigation described the procurement activity between 53 government agencies and 90 shell-companies for an amount of 26 million US dollars from 2015 to 2018.

\textsl{Case 2 – Guanajuato.}  This investigation was developed by the Citizenship Committee of the State’s Anti-Corruption System, an official agency created by local law with oversight powers \footnote{Pizano, C. Contratos a empresas fantasma en gobierno de Márquez por 192 mdp; dos están ligadas a la Estafa Maestra. Poplab (2020)}. The data collection was made by freedom of information requests. This investigation aimed to evaluate the state’s situation regarding simulated operations and to track public funding in 33 government agencies and 138 shell-companies for an amount of 23 million US dollars from 2014 to 2019.

During the periods of analysis, the states of Puebla and Guanajuato were governed by the National Action Party (PAN). In 2015, Puebla and Guanajuato represented the 5th and 6th most populated states in Mexico, respectively, each with around 5\% of the total Mexican population (approx. 119 million) \footnote{INEGI. Encuesta Intercensal 2015.}. Economically, Guanajuato represented the 6th, while Puebla the 8th, leading states in GDP with 4.5\% and 3.2\% of the total Mexican GDP in 2015 (approx. 1.172 trillion USD) \footnote{INEGI. Producto Interno Bruto por entidad federativa 2015. Boletín de Prensa Núm. 529/16.}.

\subsection*{Data}

The original reports contained data presented in various non-standard formats, which posed challenges. Mexico's subnational procurement and corporate governance transparency suffer from shortcomings as the data is often manually compiled from multiple sources, resulting in missing, incomplete, or inaccurate information. We made our best effort to build reliable and standardized datasets based on the independent original reports of the cases and complementary official sources. The sampling criteria for procurement and companies information in the original investigations was performed ad-hoc, according to the respective needs of the investigating groups and local data availability. The similarities on the amount of money involved, the number of buyers and suppliers, the temporal periods of contracting and the political party governing both states are fortunate and useful coincidences. The results derived from our datasets can only be attributed to those states and during the specified periods of time. The scope and limitations of our datasets and results must be understood within this context. Nonetheless, the datasets and analytical approach presented in this article consider the basic elements that allow researchers and analysts working on procurement or corporate ownership to apply our methodology to their respective needs.

As such, after a curation process that included cleaning and searching for additional information, our analysis considers original datasets about public procurement activities, as well as ownership and management data. The datasets about contracting activities contain information about the buyers, suppliers, spending (US dollars), contracting year, and the type of contract classified into five general groups according to the buyer's institutional sector: Government (government institutions including executive, legislative, judicial and other agencies), Education (universities and other higher education institutions), Health (public health institutions), Security (law enforcement agencies and police departments) and Social (social development and human rights agencies). This classification was chosen due to a lack of complete and consistent information about the specific type of contents of the contracts or companies' sector, therefore, our analysis regarding this aspect of the data is biased towards the sector of the buyers. However, this does not compromise the analytical rigor of the proposed methodology and general results. The data about ownership and management contain information about shareholders, administrators, legal representatives, and commissaries, as well as the date of creation (incorporation) as stated in the companies' charters. For privacy protection, we have anonymized all the information regarding the identification of government agencies, private companies and individuals. These datasets are available online (see Data Availability).

\section*{Network Analysis}

\subsubsection*{Buyer-supplier and bipartite ownership networks}

The starting point of our analysis is the buyer-supplier and ownership bipartite networks. For each case, we created unweighted bipartite buyer-supplier networks using the contracting data aggregated over the whole contracting period. The edges or links represent the existence of a contracting relation. In addition, we created unweighted bipartite supplier-individual networks (or ownership networks), that is, networks where the suppliers are connected to the individuals listed as shareholders, administrators, legal representatives, and commissaries in the companies' charters. The edges represent the existence of an ownership or management relation. For these, we computed the average degree, $\langle k\rangle$, which corresponds to the arithmetic mean over the degrees (number of neighbors), $k_i$, of all the nodes in an ordinary one-mode network or the corresponding set in a two-mode or bipartite network \cite{menczer2020}.

\subsubsection*{Connected components, diversity and favoritism}

Ownership and management data allow us to study the role and impact of connected shell-companies as extraction vehicles. For this, we considered a connected components approach. Connected components are defined as subnetworks containing one or more nodes such that there is a series of connections between any pair of these nodes but there are no connections to other components. In other words, connected components represent isolated clusters with at least one node \cite{menczer2020}. To obtain the connected components, we created unweighted one-mode suppliers networks by projecting the ownership of bipartite networks onto the set of supplier nodes. In these networks, suppliers connect with each other if they share at least one individual in a role of ownership and management in common. For each case, we identified the corresponding connected components and computed their size (number of companies conforming the cluster).

We quantify the activity diversity of the connected components based on the type of contracts obtained by their constitutive companies. The previous classification for the type of institutional buyer is used as a proxy for the type of contract (see Data). For this, we first group connected components based on their size, $n$, and make use of the Gini-Simpson diversity index,

\begin{align*}
    H_n = 1-\sum_{t \in \gamma} (p_n^t)^2
\end{align*}

\noindent where $p_n^t$ corresponds to the fraction of contracts awarded by buyer type $\gamma=$ \{Government, Education, Health, Security, Social\} to all the companies in the group of components of size $n$. The quantity $p_n^t$ is normalized per component group so that $\sum p_n^t=1$ \cite{lyra2021}. As such, $H_n=1$ is associated with the highest diversity possible, and $H_n=0$ with the lowest diversity. Thus, low/high diversity indicate different levels of specialization or generalization in procurement activities. In particular, a component group with low diversity might indicate potential coordination or cooperation of the constitutive companies in order to control a specific buyer or region of the market. Also, notice that although this index is applied to groups of connected components, it can also be directly applied to the connected components themselves and even single entities \cite{lyra2021}.

We quantify the favoritism of the suppliers by a particular buyer based on the total number and spending acquired through contracts. For this, we make use of the weighted favoritism coefficient \cite{IMCO2018,falcon2022practices} given by,

\begin{align*}
    F_{ij} = \frac{1}{3}\biggl(\frac{p_{ij}}{p_j}\biggr) + \frac{2}{3}\biggl(\frac{w_{ij}}{w_j}\biggr)
\end{align*}

\noindent where $p_{ij}$($w_{ij}$) represents all the contracts(spending) between a supplying entity $i$ and a buying entity $j$, while $p_j$($w_j$) is the total number of contracts(spending) of the buying entity $j$ in the same period of analysis. Notice that this coefficient can be reduced to represent the favoritism of a single supplying entity by adding the contributions of all buying entities in the market, that is, $ F_i = \sum_j F_{ij} = (1/3)p_i + (2/3)w_i$, where $p_i$($w_i$) is now the fraction of contracts(spending) of the supplying entity $i$ across all the market during the period of analysis. As such, $F_{ij}=1$ is associated with the highest favoritism, and $F_{ij}\to 0$ with low favoritism \cite{IMCO2018,falcon2022practices}. Also, notice that the supplying entity $i$ can be a single company or group of companies, and likewise, the buying entity $j$ can be a buyer or group of buyers. Here, the favoritism coefficient is applied to groups of connected components across all the market as well as specifically among groups of buyers or modules (as defined below).

\subsubsection*{Similarity and module-component representation}

In a further step to understand the role and impact of the shell-company networked operations, we developed a module-component (MQ) bipartite representation of the procurement markets based on the structural similarity of the buyers co-contracting networks. First, we created buyer-component bipartite networks where buyers are connected to the shell-company clusters (connected components) grouped by their size $n$. Then, we created one-mode weighted buyers networks by projecting the buyer-component bipartite networks onto the set of buyers nodes. The weight of the connections is based on the similarity of co-contracting activity and measured by the Jaccard similarity coefficient,

\begin{align*}
    w_{ij} = \frac{|\eta_i \cap \eta_j|}{|\eta_i \cup \eta_j|},
\end{align*}

\noindent where $\eta_i$($\eta_j$) is the set of neighbors of node $i$($j$), and the vertical bars indicate the cardinality of the set \cite{Wachs2019}. We identify the modules, $M$, using the Louvain community detection algorithm \cite{Blondel_2008}. For our analysis, the chosen partition is the one with the highest probability of occurrence after 1000 randomizations of the node and the community evaluation order in the Louvain algorithm. A module or community is defined as a set of nodes that are relatively more tightly connected among them than with the rest of the nodes in the network \cite{menczer2020}. As such, these modules represent connected regions within the buyers networks (or market) defined by their co-contracting activity patterns. We also quantify the diversity of these modules based on the institutional type of the constitutive buyers and the Gini-Simpson diversity index.

The network analysis was performed using NetworkX \cite{hagberg2008exploring} and custom Python code. The network visualizations were created using Cytoscape \cite{shannon2003cytoscape}.

\section*{Results}

\begin{figure}[t!]
\centering
\includegraphics[width=1.0\textwidth]{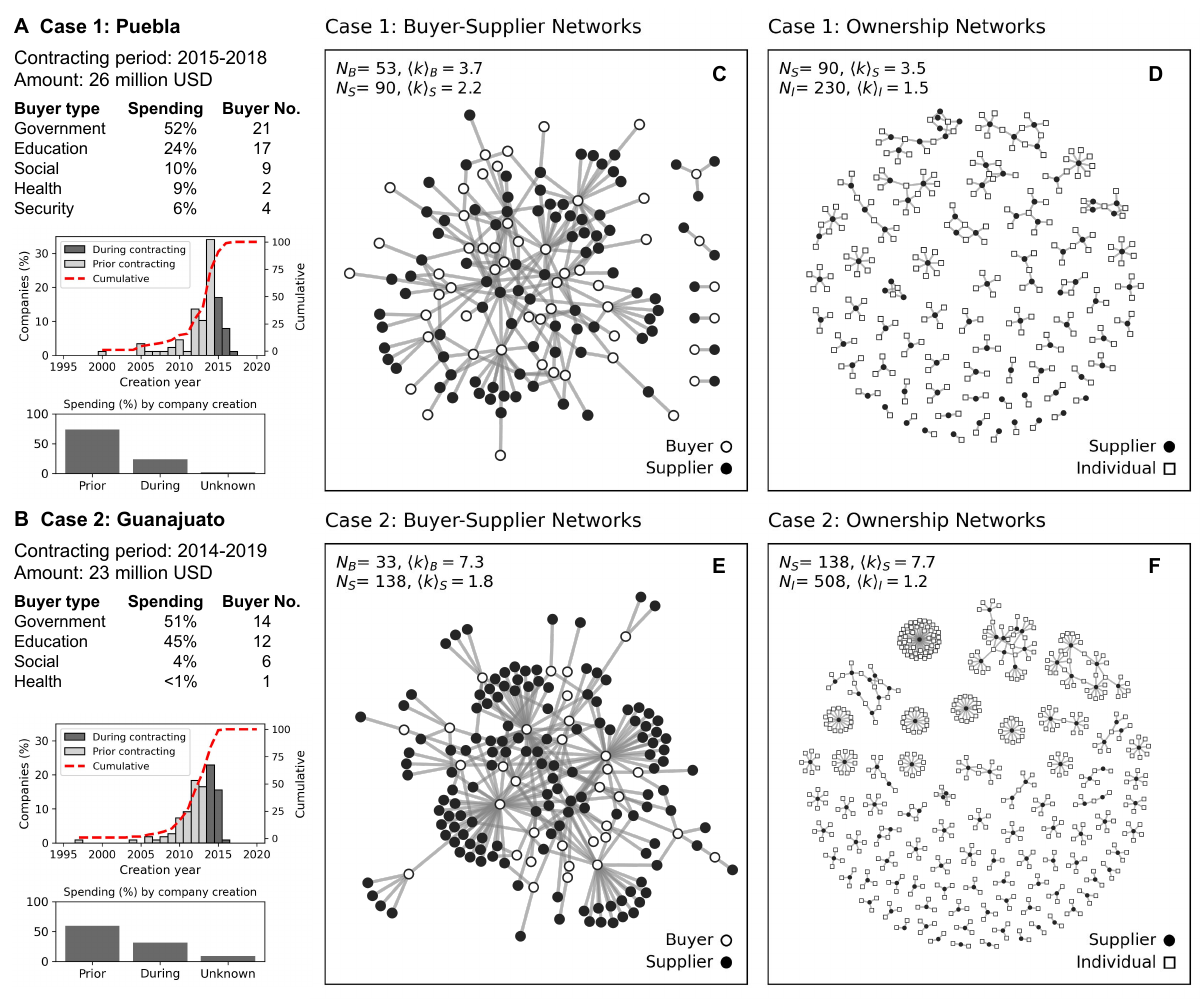}
\caption{\small \textbf{Cases and bipartite network visualizations.} In (A) and (B), general features of the cases, including: the spending distribution according to buyers type, the number of companies created prior to the contracting period, and the spending obtained by these companies. The ``Unknown'' label corresponds to shell-companies with missing information. In (C) and (E), the buyer-supplier networks of each case with the number of buyers, $N_B$, and suppliers, $N_S$, as well as their corresponding average degree, $\langle k \rangle$, are indicated. In (D) and (F), the ownership networks with the number of suppliers, $N_S$, and individuals, $N_I$, as well as their corresponding average degree, are indicated.}
\label{fig:fig1}
\end{figure}

Our comparative analysis starts by creating buyer-supplier and supplier-individual (ownership) networks, as well as by establishing some general characteristics for each case (Fig. \ref{fig:fig1}). As previously stated, the states of Puebla and Guanajuato share similar social, economic, and political characteristics around the period of analysis despite not being geographic neighbors (see Cases). Similarities are also found in the characteristics of the procurement activities and shell-companies involved (see Figs. \ref{fig:fig1}A and \ref{fig:fig1}B): (\textit{i}) the spending percentage extracted was higher ($>50\%$) in buyers within the Government and Education categories, (\textit{ii}) most shell-companies ($>50\%$) were created prior to the corresponding contracting period, and (\textit{iii}) these companies created prior to the contracting activities were those that extracted more of public funds ($>50\%$). Recall, that incorporation or creation dates of the companies were obtained directly from the companies’ charters with the only criterion that these companies were contracted during the corresponding periods of each case (Puebla from 2015 to 2018, and Guanajuato, from 2014-2019), and not from other previous ad-hoc temporal periods. Regarding the buyer-supplier networks (Figs. \ref{fig:fig1}C and \ref{fig:fig1}E), we found that they tend to form one big connected component, typical of high concentrated markets \cite{Wachs2020}. On the other side, the structure of ownership networks is sparse (Fig. \ref{fig:fig1}D and \ref{fig:fig1}F), tending to form many connected components due to the existence of shared individuals \cite{nicolas2021}. 

\begin{figure}[t!]
\centering
\includegraphics[width=1.0\textwidth]{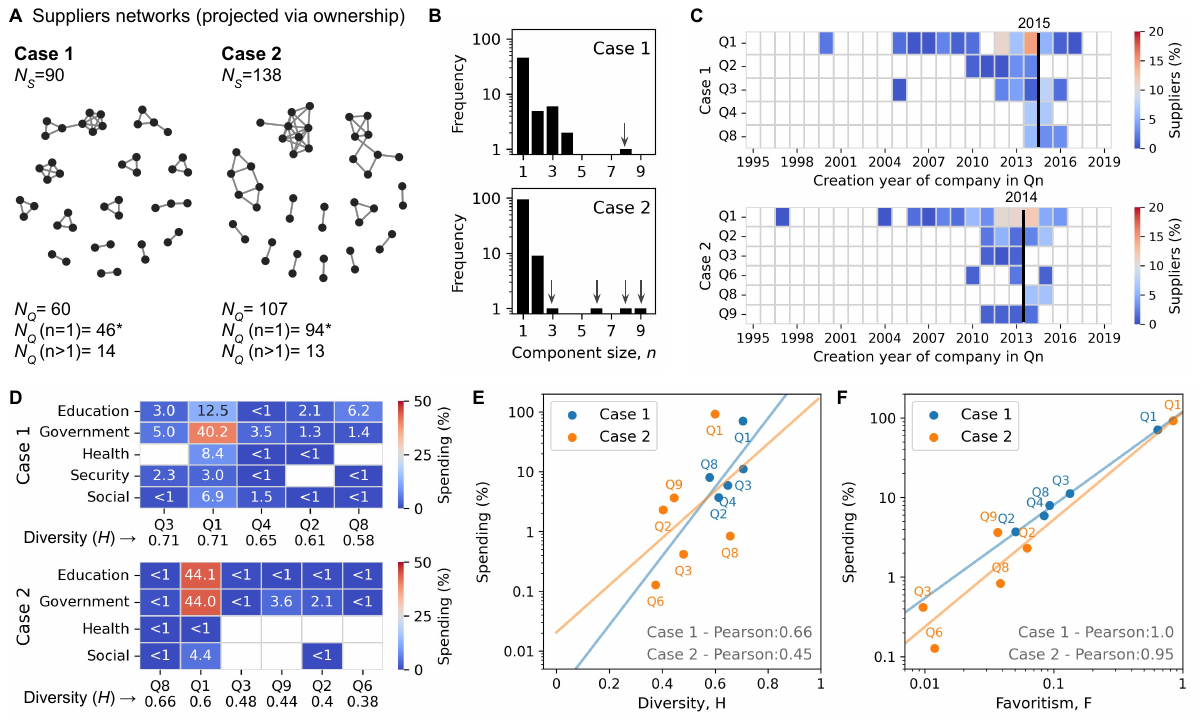}
\caption{\small \textbf{Connected components (shell-company clusters).} (A) Connected components, $Q$, in the one-mode suppliers networks obtained via projection of the bipartite ownership networks. The number of single ($n=1$) and multi node components ($n>1$) are indicated. Single-node components are not shown for visualization purposes. (B) Distribution of the connected components according to their size, $n$, in a semi-log plot. Arrows indicate unique components of size greater than one. (C) Temporal distribution of the companies according to the size of the component they belong to (labeled as ``Qn''). The dark solid lines indicate the beginning of the corresponding contracting periods. (D) Spending per component size and contract type. The diversity index, $H$ is also indicated. In (E) and (F), the diversity index and favoritism coefficient versus the spending percentage per component size in semi-log and log-log plots, respectively. The Pearson correlation coefficient of the linear fits is indicated.}
\label{fig:fig2}
\end{figure}

In a first approach to study the role and impact of these shell-companies, we created one-mode suppliers networks by projecting the ownership bipartite networks unto the set of companies (Fig. \ref{fig:fig1}D and \ref{fig:fig1}F). This projection creates networks in which the connected components, $Q$, are clusters of shell-companies of size, $n$ (see Fig. \ref{fig:fig2}A and \ref{fig:fig2}B). Recall that the minimum size of a connected component is one (see Network Analysis). By forming groups of connected components according to their size, we found that: most of the shell-companies that constitute these clusters were created a few years before starting the corresponding contracting periods (2015-2018 and 2014-2019, for case 1 and 2, respectively); Q1 components were created even a decade before (the oldest ones created in the 90's) and with a higher frequency just one year before the starting of the contracting periods, while components of greater size ($n>3$) tend be created around this period (see Fig. \ref{fig:fig2}C). This suggest that these shell-companies were operating before the current period of analysis and that their real economic impact could be even higher than what is reported in the investigations.

The contracting activities diversity per component size (Fig. \ref{fig:fig2}D) given by the Gini-Simpson index, $H$, shows that Q1 components not only have a high diversification, but also the highest impact on diverted funds, being Education (public universities and other higher education institutions) and Government (government institutions including executive, legislative, judicial and other agencies) the sectors with most damages, approximately $75\%$ and $95\%$ of the total embezzled funds of Case 1 and 2, respectively. Furthermore, although we did not find strong evidence of a positive correlation between the amount diverted and the components' diversity (Fig. \ref{fig:fig2}E), we found that the amount diverted was strongly correlated with the overall components' favoritism, $F$, that is, by considering the total spending contributions of all buyers.

\begin{figure}[t!]
\centering
\includegraphics[width=1.0\textwidth]{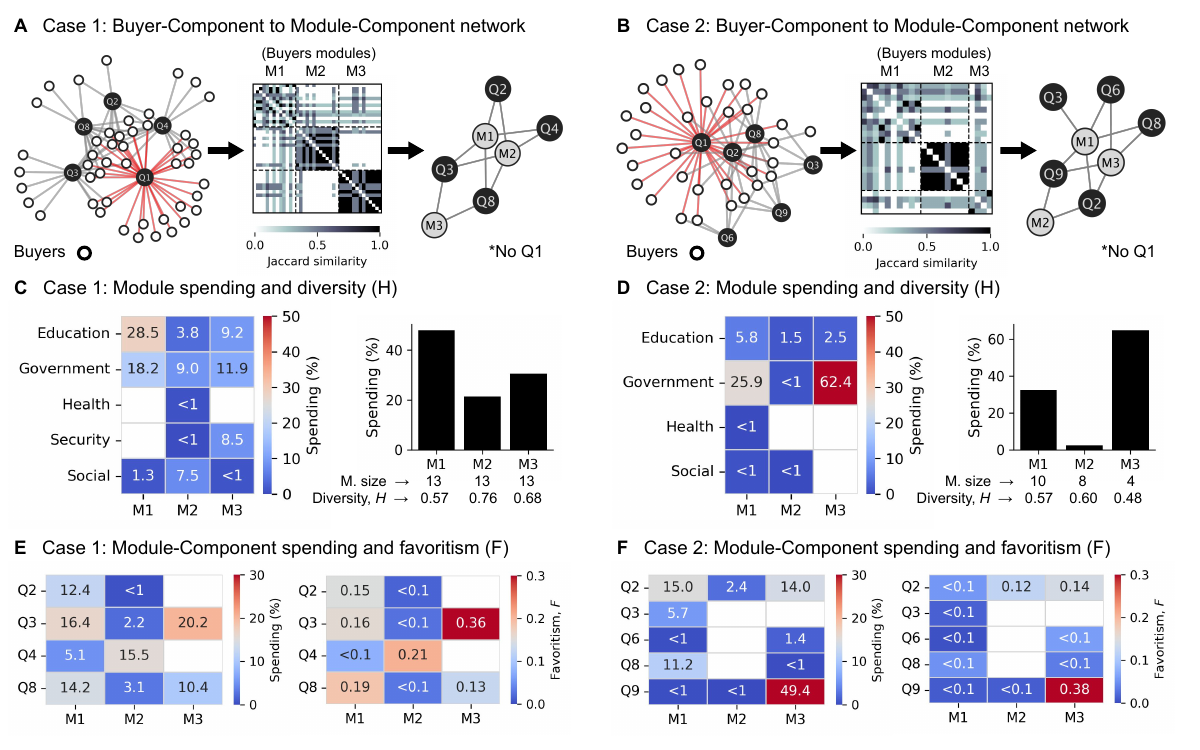}
\caption{\small \textbf{Module-component networks.} In (A) and (B), illustrations of the buyer-component to module-component bipartite network transformation. Connected components are grouped according to their size and labeled as ``Qn''. The modules, $M$, are detected in the projected buyers network (shown as an adjacency matrix visualization) and labeled M1, M2 and M3, respectively. For analytical purposes, the Q1 node was removed before projections (see main text). In (C) and (D), distribution of the spending per module and contract type, with the module size and diversity index indicated. In (E) and (F), visualizations of the module-component networks, shown as adjacency matrices, with the favoritism coefficient and diversity index as indicated.}
\label{fig:fig3}
\end{figure}

In a further step to understand the role and impact of the shell-company networked operations, we developed an alternative module-component bipartite approach. In this approach, the set of buyers are partitioned into different groups or modules, $M$, according to their co-contracting patterns of shell-company clusters. To do this, we first create buyer-component bipartite networks where buyers are connected to the shell-company clusters (connected components) grouped again by their size (see Fig. \ref{fig:fig3}A and \ref{fig:fig3}B). Second, we create weighted buyer networks by projecting the buyer-component bipartite network into the set of buyers using the Jaccard similarity coefficient as edge weights. Finally, we apply the Louvain method of community detection to group the buyers into modules according to their co-contracting similarity (see Network Analysis). In addition, in order to gain a better understanding of the role and impact of shell-company clusters ($n>1$), all the information of Q1 components are removed for this part of the analysis.

By applying the module-component transformation, we found that in both cases the complexity of the markets is reduced to just three buyers modules, labeled as M1, M2 and M3, respectively (Fig. \ref{fig:fig3}A and \ref{fig:fig3}B). These modules are constituted by buyers of different types (see Fig. \ref{fig:fig3}C and \ref{fig:fig3}D): in Case 1, all modules have the same size and concentrate most of the spending in the Education and Government categories, while in Case 2, modules have different sizes and most of the spending is concentrated in the Government category. In both cases, modules exhibit similar diversity but are impacted differently in monetary terms, especially those with the lowest diversity. In addition, we found that in the MQ network representation (presented in Fig. \ref{fig:fig3}E and \ref{fig:fig3}F as adjacency matrices), there is an agreement between the module-component connection with the highest favoritism measured and the highest spending, which for Case 1 is the connection of M3 with the Q3 components with approx. 20\% of the total spending, while for Case 2 the connection of M3 (the smallest module with the lowest diversity) with the Q9 component (the largest shell-company network) with approx. 50\% of the spending. Recall that the contribution of Q1 components was removed from the spending percentage.

\section*{Discussion}

\subsection*{The module-component framework}

The high level of connectivity and the emergence of one connected component in both the buyer-supplier and module-component network representations indicate that the contracting of shell-companies was not an isolated phenomenon of just a few public institutions but instead, a generalized mechanism of extraction across the whole procurement ecosystem regardless of the sector of the buyers, favoritism patterns or the spending extracted (see Fig. \ref{fig:fig1} and \ref{fig:fig3}). 

Notably, we found that single shell-companies have the most economic impact overall (see Fig. \ref{fig:fig2}D). Although many of the shell-companies forming connected components of size greater than one were created a few years prior the contracting activity of the cases, it's unclear that this specific scheme in itself was used to maximize profits, that is, there's not a clear positive correlation between the connected components size and the spending extracted (see Fig. \ref{fig:fig2}E and \ref{fig:fig2}F). In addition, the similar temporal patterns of companies incorporation and, in particular, the peaks observed before/after the start of the contracting periods (see Fig. \ref{fig:fig1}A-B) seem to be features of the sampling method of the contracts and not attributable to any other economic or political factors. Note that the sampling of the contracts in the original journalistic investigations was merely to find illicit and criminal activities and not necessarily for formal technical analysis. However, there is a contextual element that could potentially explain both peaks observed. From 2011 to 2017, and from 2012 to 2018, the states of Puebla and Guanajuato, respectively, were under fixed (political) administrations, establishing relevant windows of action for opportunistic groups with particularistic ties or relevant resources to tap into the local procurement markets. These fixed time periods put the peaks of company creation just at the middle of both local administrations (around 2014-2015) signaling the maximum level of coordinated actions. Further evidence is required to formally prove such a hypothesis.

Furthermore, the lack of strong evidence for a clear positive correlation between the connected components size and the spending extracted could be accounted by a few factors. First, the life cycle of a company that is meant to become catalogued as shell by the fiscal authorities is intermittent and finite. Shell companies are active or inactive over the years until they are awarded contracts. Some were originally conformed to remain legal for the purpose of camouflaging the operations while others to remain illegal to hide the beneficiary ownership and be dissolved when public funds are transferred. The latter is because shell companies are often used for tax avoidance in Mexico \cite{Zumaya2021}. In the procurement context, they are dissolved after turning the diverted money into non-traceable cash. This was documented in both cases as this is a typical behavior identified in the country that facilitates the concealment of the beneficial ownership and entangles accountability of funds and expenditure audits \cite{deacha2018empresas}. Second, from a heuristic perspective, it could be considered that networked companies usually pose a greater risk for criminal activity given that untangling the shell-network and operational schemes could be done relatively more easily once the companies' ownership is known. In Mexico, this has proven to be right in multiple investigations regarding shell-companies and corruption \cite{Luna-Pla2020, nicolas2021}. Third, a strong positive correlation between size and amount extracted is not universal among procurement corruption cases; in such cases, further evidence should be gathered. 

Overall, our results not only signal potential fraudulent operations of economic cartels that have operated for many years but also a form of organized crime emerging from the group activities. Indeed, the integral analysis of procurement and ownership data provides further elements to the description and detection of cartel activity and extra-legal governance through procurement \cite{fazekas2022extra, adam2022public}. Although these approaches also consider the risk brought by particularistic ties among political, business and criminal groups, these dark ties are not considered explicitly due to their nature \cite{bond2011political}, but it does not mean that they could not exist. The misuse of companies is often a vehicle to funnel illegal money into political campaigns and vote buying, by issuing fake invoices to governments as a method to divert funds. In the Puebla and Guanajuato cases, electoral changes occurred during the contracting periods in 2017 and 2018, respectively. In addition, withholding political and economic power is a dynamic nature of legal and illegal companies acting as multiple networks, as it allows the political individuals or criminals infiltrated in the group to remain across time. This property of organized crime seems to be effective in enduring changes of political groups and retaining the control of procurement market territories \cite{reeves2017bid, ribeiro2018dynamical, martins2022universality, waxenecker2019impunidad, gambetta1993sicilian}. However, the evidence to explicitly show or prove that such coalition among politicians, government officials and economic groups existed falls beyond the scope of our current results.

\subsection*{Insights for preventing organized crime in procurement}

Organized crime literature pointed out how to address the dark ties in procurement environments from the perspective of a company's behavior acting as a group in international and domestic corruption cases \cite{gambetta1993sicilian, jancsics2017offshoring}. Such behavior is regarded to benefit an organization and is often conceptualized as illegal corporate behavior and economic, corporate or business crime because of the complexity of the relationships and communications \cite{VanDuyne2006, pinto2008corrupt, slingerland2018network, Slingerland2021}. In networks characterized by the misuse of companies in procurement, companies are not organizations of crime as such, but rather the method with which powerful groups perpetuate activities and economic enterprise as a vehicle to be used according to convenience \cite{deWillebois2011}. Within this conceptual framework, collaboration takes shape through concealed connections formed by means of bribes and extortion payments exchanged between companies and public officials, or vice versa. These illicit transactions, which form an integral part of the business model, are orchestrated to secure contracts and purchases. Various individuals within the network, including judges, prosecuting officials, notaries, and tax auditors, derive personal benefits from their involvement, where the beneficial ownership is deliberately obscured within the intricate structure of the network \cite{nielsen2003corruption, Campbell2018}. 

Because much of the crime committed by private corporations, politicians, and government agencies in the procurement environment is deeply harmful, greater precision is needed to understand the differences in the interrelated behavior of suppliers and buyers, in order to be anticipated for prevention \cite{Albanese2018, fazekas2022extra, falcon2022practices}. Our research showed the shell companies' organizational patterns can be described as networks of connected actors that exhibit consistent growth and development in comparable situations. This aligns with the scholarly portrayal of these networks as opportunistic collaborations among individuals aimed at achieving specific objectives and purposes. Such networks are often regarded as organized crime entities, and differ from traditional mafia groups in terms of their organizational structure, recruitment strategies, and governance characteristics, as documented in current literature \cite{Campbell2018, ReuterandTonry2020, ReuterandPaoli2020}. Organized criminal studies address the changing nature of the entrepreneur's criminal environments and the levels of analysis, from individuals, groups, and the legal-illegal nexus \cite{VanDuyne2006, vonLampe2006}. Entities have the potential to transition from conventional forms of commercial or public corruption to organized crime. This progression could take place gradually, starting with instances of bribery and white-collar crime, eventually transforming into a fully operational criminal enterprise employing tactics like fraud, solicitation, and extortion \cite{Albanese2018}. Among many other factors, this changing nature of the phenomena blurs the conceptualization of organized crime and makes it complex to measure.

This paper contributes to crime prevention by providing quantitative means of examining the economic ramifications of isolated versus interconnected shell company operations, and incorporating diversity and favoritism metrics to improve the precision of proactive measures. Hence, the analytical method presented in this paper can be applied to other procurement datasets or various other manifestations of corruption that may escalate into organized crime, for instance, to an agency afflicted not merely by a few corrupt officers but by systemic corruption \cite{Albanese2018}. In general, given the inherent spontaneity and non-scripted nature of social interactions in real life, the utilization of predictive technology is contingent upon the availability and reliability of data and the application of computer science techniques to quantify relevant phenomena and environmental factors \cite{Edwards2017}. As such, data analysis and computer science offer substantial potential for prediction by effectively analyzing information while simultaneously defining the organizational type and structure of the network under examination. In particular, network analysis serves as an additional indispensable tool in the red flag process of risk assessment \cite{Wachs2020, adam2022public, Kertesz2020}, working in conjunction with forensic investigations to facilitate a comprehensive understanding of the collaboration dynamics, the individual roles within the network, the nature of criminal activities being carried out, and the boundaries that define the network's scope \cite{Bouchard2020}.

A systematic approach also allows data-driven solutions that stem from changing regulations (\cite{Albanese2018, Luna-Pla2020}, such as raising the difficulty of creating companies and restricting the types of products and services they can offer, enhancing the person identification rules, increasing powers of enforcement and surveillance; legislating, increasing transparency and oversight to the registration of ownership and directorship of companies \cite{Campbell2018}. New forms of control, such as taxing the proceeds of crime and income tax to natural or legal persons, can be successful in prosecuting when other organized crime conducts are difficult to prove \cite{Friel2018}, for example, crimes against the state property, when public servants and politicians are entrenched, leaving no incentives on the government side to act in the interest of the public good \cite{nielsen2003corruption, persson2010failure}. Due to the limitations of existing measures in various environments, the proposed approach aims to identify the operational priorities regarding legal structures and jurisdictions, potentially redefining the governability of the business sector by introducing innovative forms of cooperative action \cite{Campbell2018}. Its objective is to address the issue of criminalizing corporate misconduct, considering that corporations can potentially provide opportunities for crime. Additionally, this approach seeks to gain a deeper understanding of the evolving properties of networks within dynamic contexts. These changing circumstances require a fresh perspective on corporate liability and the effectiveness of preventive measures \cite{Cronin2018}.

\section*{Final remarks}

The cases studied here show that corruption networks in procurement markets vary in structure and multiple illegal activities that unravel an organized crime \textit{modus operandi}. Therefore, focusing only on either the contracting relations or ownership structures in isolation, without addressing the multilayer nature of procurement markets in their political and power context, may significantly limit the impact of prevention and prosecution efforts \cite{jancsics2017offshoring}. Further research should look at the challenges to find empirical evidence on the groups' activities that maximize profit, money tracking information and beneficiary ownership identification \cite{Diepenmaat2021}. In order to effectively prevent corruption, it is crucial to develop tools that can assess the progression from less to more corrupt behaviors among both private and public individuals. By understanding this progression, it becomes possible to intervene and disrupt the factors, both legal and economic, that enable and encourage the escalation of this type of organized crime.

Finally, there is a tendency to believe that corruption developed in a group scheme is organically organized, coordinated at all levels in a hierarchical way, and run automatically \cite{nielsen2003corruption}. While this could be the case in some networks, organized crime networks are better approached by understanding the nature of the groups, that is, the diversity of their cooperation within a system and context, the level of influence and links among members \cite{felson2006ecosystem, carrington2011crime, Luna-Pla2020}. When groups are informal, the beneficial ownership is hidden, information is scarce, and the end purposes are unclear, network approaches helps to depict the complexity of the problem by offering a global view of the activities perpetuating the illicit business over time \cite{morselli2013crime, Wachs2020, ribeiro2018dynamical, martins2022universality}, as well as intervention and control strategies according to specific environments \cite{von2015organized, da2018topology, Solimine2021}.


\section*{Data Availability}

The data used in this article is available from the following repository: \url{https://doi.org/10.6084/m9.figshare.21902160}


\bibliographystyle{apalike}
\bibliography{references}


\section*{Conflict of interests}

The authors declare that the research was conducted in the absence of any commercial or financial relationships that could be construed as a potential conflict of interest.

\section*{Authors contributions}

Both authors conceived and developed the research idea. ILP developed the criminal theoretical analysis. JRNC performed the computational analysis and made the figures. Both authors participated in the discussion, writing and approval of the final manuscript.

\section*{Acknowledgments}

ILP acknowledges the notaries, legal experts, administrative judges, and high-level public servants in the Mexican Financial Intelligence Unit, for the interviews that took place from January 2021 to March 2022, with the aim of gaining a deeper understanding of the legal, political, and criminal context of the cases studied. ILP also acknowledges the support of PASPA-DGAPA at UNAM for providing funds for the research of this paper.

\end{document}